\documentclass[a4paper,superscriptaddress,floatfix,showkeys,twocolumn,prl]{revtex4}
\usepackage{graphicx}
\begin{document}

\title{Richardson's pair diffusion and the stagnation point structure
  of turbulence}

\author{J. D\'avila}
\email{davila@eurus2.us.es}
\affiliation{E. Superior de Ingenieros, Camino de los Descubrimientos s/n, 
41092 - Sevilla, Spain}
\author{J.C. Vassilicos}
\affiliation{Department of Aeronautics, Imperial College, London SW7 2BY, UK}

\date{\today}
\begin{abstract}

DNS and laboratory experiments show that the spatial distribution of
straining stagnation points in homogeneous isotropic 3D turbulence has
a fractal structure with dimension $D_s = 2$. In Kinematic Simulations
the time exponent $\gamma$ in Richardson's law and the fractal
dimension $D_s$ are related by $\gamma = 6/D_s$.  The Richardson
constant is found to be an increasing function of the number of
straining stagnation points in agreement with pair
duffusion occuring in bursts when pairs meet such points in the flow.

\end{abstract}

\maketitle

The rate with which pairs of points separate in phase space or in
physical space is of central importance to the study of dynamical
systems. Pairs of points in the phase space of a low-dimensional chaotic
dynamical system separate exponentially. In chaotic advection pairs
also separate exponentially \cite{O89} leading to exponentially
fast stirring and a high potential for mixing.  In turbulent flows,
however, pairs of fluid elements separate on average algebraically
\cite{O41,B50,OM00}. Turbulent flows have a very wide range of excited
length- and time-scales and are therefore fundamentally different from
both low-dimensional dynamical systems and chaotic advection flows.
Attempts have been made to model the relative diffusion of fluid element
pairs in terms of Langevin type equations based on the assumption that
relative Lagrangian accelerations are Markovian in time \cite{PN94,H97}.
These models can reproduce the right algebraic time growth of separation
statistics of fluid elements in turbulent flows; but they fail to explain
the very large values taken by the flatness factor of Lagrangian relative
velocities in Direct Numerical Simulations of isotropic turbulence
\cite{Y94,H97,MV99}. In fact these models based on Markovian acceleration
statistics underestimate this flatness factor by as much as one order of
magnitude.

The algebraic growth of Lagrangian separation statistics can also be
accurately reproduced by Kinematic Simulations \cite{FHMP92,EM96,FV98}.
These are models of turbulent diffusion based on kinematically simulated
turbulent velocity fields which are non-Markovian (not delta-correlated
in time), incompressible and consistent with up to second order statistics
of the turbulence such as energy spectra.  Kinematic Simulations are
interesting in particular because they do reproduce the very high flatness
factors of Lagrangian relative velocities \cite{MV99}.
The mechanism by which fluid element pairs separate in Kinematic
Simulations (KS) might therefore be comparable to the one in turbulent
flows and is clearly different from the Wiener process which causes
fluid element pairs to separate in Lagrangian models of relative
diffusion based on Langevin type equations. But what is this mechanism
and why can it give rise to the algebraic growth of relative separations?

Richardson's law of turbulent relative diffusion states that the mean
square distance between two fluid elements $\overline{\Delta^{^2}} $
is proportional to the third power of time $t$, i.e.
\begin{equation}
  \overline{\Delta^{^2}}(t) \approx
        G L^{2} \left(\frac{t u'}{L}\right)^{\gamma},
  \label{Richardson}
\end{equation}
where $ \gamma =3 $, $G$ is a universal dimensionless constant and
$L$ and $u'$ are the integral length-scale and the rms velocity of
the turbulence respectively. Richardson's law is expected to be valid
in homogeneous isotropic turbulence and for times $t$ such that $
\overline{\Delta^{^2}} $ is within the inertial range of scales, i.e.
$ \eta^2 \ll \overline{\Delta^{^2}} \ll L^2 $, where $\eta$ is the
Kolmogorov microscale. Fluid element pairs follow close trajectories
for long stretches of time and separate violently when they meet
straining flow regions around stagnation points (straining stagnation
points) \cite{FHMP92,FV98,JPT99,NV02}. \cite{MV99} noted that the very
high flatness factors of Lagrangian relative velocities are consistent
with this conjecture.

\cite{FV98} found evidence of a fractal spatial distribution of
straining flow regions in KS homogeneous isotropic turbulent velocity
fields. In this paper we quantify their observation by showing that
the number of straining stagnation points per unit volume
in KS, Direct Numerical Simulations (DNS) and laboratory experiments 
of homogeneous isotropic turbulence is given by
\begin{equation} 
   n_s \approx C_s L^{-3} \left(\frac{L}{\eta}\right)^{D_s} 
\label{Davilicos}
\end{equation} 
where $ D_s =2  $ and $C_s$ is a dimensionless number. The exponent $D_s$
can be interpreted as a fractal dimension.  Both (\ref{Richardson})
and (\ref{Davilicos}) hold for $L/\eta \gg 1$.

We now describe the DNS and the grid turbulence measurements used to
establish (\ref{Davilicos}) with $D_s$=2. We have used DNS data of
non-decaying homogeneous isotropic incompressible turbulence generated 
by a standard pseudo-spectral 
code with grid resolution of about $2\eta$ and have computed the number 
of stagnation points in instantaneous 
velocity fields ${\bf u} = {\bf u}({\bf x}) = (u({\bf x}), v({\bf x}), 
w({\bf x}))$ for a variety of Taylor microscale Reynolds 
numbers $Re_{\lambda}$ ranging from 34 to 130. In this paper we 
focus our interest on the straining flow regions around stagnation 
points of ${\bf u}$. A stagnation point does indeed correspond to a 
straining flow region when the eigenvalues of the velocity gradient 
matrix at this point are all non-zero (see Ottino 1989); these are 
regions where the flow is always straining and may or may not be rotating 
as well. 

\begin{figure}
\includegraphics[width=8cm]{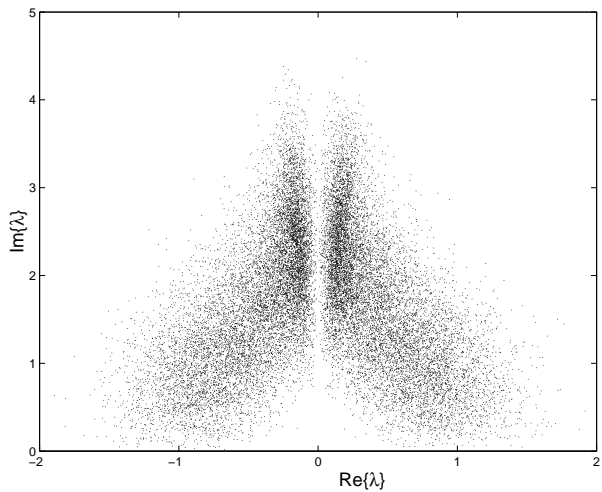}
\caption{Imaginary versus real part of the complex eigenvalues of the
   gradient of velocity matrix with positive imaginary part for 26726
   stagnation points in a KS field with $L/\eta\approx 45$.  Very few
	complex eigenvalues  with small (absolute) real part can be found.
	Similar results (with poorer resolutions) were obtained with DNS for
	similar values of $L/\eta$.\label{fig1}}
\end{figure}

\begin{figure}
\includegraphics[width=8cm]{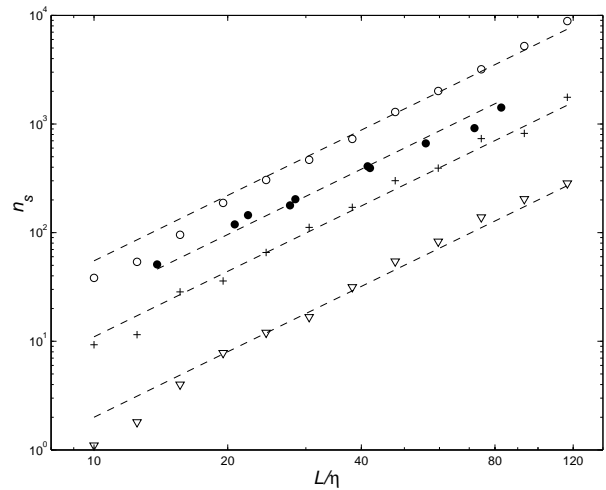}
\caption{Number of stagnation points per unit volume versus
         $L/\eta$ in DNS ($\bullet$) and KS for $p=5/3$ and different
	 values of V (compared to $u'=1$): $\circ$ $V\!\!=0$,
         + $V\!\!=0.7$, $^\bigtriangledown$
	 $V\!\!=1.0$. Dashed lines representing $ n_s \propto
	 (L/\eta)^2 $ are shown for comparison. \label{ns_KS}}
\end{figure}

We use the Newton-Raphson method (tested against the amoeba method
in various planar flows) to find all stagnation points \cite{Pet92}.
This is an iterative method and requires starting points, which have
been taken all over the DNS field at points separated by a distance
smaller than $\eta$. Irrespective of Reynolds number, the vast majority
of stagnation points have turned out to be in straining regions. Figure
\ref{fig1} is a scatter plot of the complex eigenvalues of the gradient
of velocity matrix. Notice that the probability to find eigenvalues
with imaginary part much larger than the real part is very low. The
only case where stagnation points can be non-straining is when two
eigenvalues are pure imaginary and the third eigenvalue is zero.
Hence the number of stagnation points is therefore effectively the same
as that of straining stagnation points, and this number per unit volume
is $n_s$. For every Reynolds number, we calculate $n_s$, $L$ and $\eta$
and we plot $n_s$ as a function of $L/\eta$ (see figure \ref{ns_KS}).
The relation between these two quantities appears to be well fitted by
(\ref{Davilicos}) with $D_s=2$.

Experimental support for (\ref{Davilicos}) with $D_s = 2$ has been 
obtained from grid generated turbulence in the laboratory. The
turbulence is homogeneous and isotropic far enough from the grid
and from the wind tunnel walls. Measurements of the streamwise
air velocity $u$ were taken with a hot wire at the centre of the working 
section of a wind tunnel at a distance of 50 mesh sizes behind the grid. 
We collected thirty runs of data for thirty different values of
$Re_{\lambda}$ ranging from 68 to 130. The sampling frequency was 30\,kHz,
enough to resolve the dissipation range except at the highest values of
$Re_{\lambda}$ but always enough, however, to resolve the Taylor microscale 
in all our runs. Each data set contains more than 100 integral scales. 
The turbulence intensity was always smaller than 5\% thus allowing the 
use of the Taylor hypothesis for the conversion of temporal data into
spatial data.  It is possible, from these wind tunnel data, to calculate
the number of zero-crossings of $u$ per unit length which leaves us with
the problem of relating this number to the number of stagnation points
per unit volume in homogeneous and isotropic turbulence. Each component
of ${\bf u}$ has an instantaneous zero-crossing surface in the
three-dimensional space of the flow. These three instantaneous surfaces
may have a fractal dimension larger than or equal to 2. Because of isotropy, 
these three fractal dimensions must be the same and we denote them by $D$. 
Stagnation points of ${\bf u}$ are intersections of the three zero-crossing
surfaces. The rule of the thumb for estimating the fractal dimension of
intersections of surfaces is that the codimension is equal to the sum of
the codimensions of the intersecting surfaces \cite{F90}. The codimension
of each zero-crossing surface is $D-2$ because surfaces are two-dimensional
and the codimension of the set of their intersections is $D_s -0$ because 
points are zero-dimensional. Hence, $D_s = 3(D-2)$. 
We are assuming our DNS observation that the vast 
majority of stagnation points lie in straining regions to be also true in 
grid generated turbulence which is why we use the notation $D_s$ for the 
fractal dimension of the stagnation points of ${\bf u}$. 

By virtue of the Taylor hypothesis, the zero-crossings of our
one-dimensional $u$ data can be viewed as a set of point intersections
through the zero-crossing surface $u({\bf x})=0$. From our one-dimensional
data the codimension $D-2$ can be measured from one data set with a specific
Reynolds number by applying a box-counting algorithm to the zero-crossings.
This method was used by Sreenivasan and his colleagues (see \cite{S91}
and references therein) who found that the fractal codimension $D-2$ is
indeed well-defined and equal to $2/3$.
Another way to measure $D-2$ is to count the number of
zero-crossings in different data sets corresponding to different
Reynolds numbers and relate the zero-crossing numbers per unit length
to $L/\eta$. However, because we do not resolve $\eta$ in some of our 
runs, we apply a low-pass filter of wavenumber $2\pi/l_c$ in order to
remove the poorly resolved dissipation range and the high-frequency
electronic noise of our measuring device.
The filter wavenumber is in fact equal to $2\pi/\eta$ for the smallest
Reynolds numbers and turns out to be closer to the Taylor microscale
wavenumber in most cases. This is the method we have applied here and
we have found that the zero-crossing number per unit length is
proportional to $(L/l_c)^{2/3}$ which implies $D-2 = 2/3$. From 
$D_s = 3(D-2)$ we deduce $D_s =2$ in support of our DNS findings. 

To explore the relation between the turbulent diffusion of fluid element
pairs and the fractal structure of straining stagnation points in the
flow, i.e.  between (\ref{Richardson}) and (\ref{Davilicos}), we need
to find ways to modify the spatial distribution of straining stagnation
points and monitor the changes in turbulent pair diffusion brought about
by such modifications.  Such a study cannot  be carried out with current
DNS and laboratory experiments of homogeneous isotropic incompressible
turbulence because the spatial distribution of straining stagnation
points is determined by the Navier-Stokes dynamics and cannot be tampered
with.  KS, however, offers the flexibility to chose the energy spectrum
at will and thereby modify, as  we show below, the fractal structure of
the set of straining stagnation points.  An additional advantage of KS
is that the Lagrangian pair diffusion statistics it produces compare well
with DNS results when the energy spectrum chosen is that of the DNS
turbulence \cite{MV99}. KS also succesfully generates \cite{NV02} all the 
pair diffusion results of the laboratory experiment of \cite{JPT99}.

 KS uses turbulent-like velocity fields of the form
\vspace*{-7mm}

\begin{eqnarray}
   {\bf u} = \sum_{n=1}^{N_k}
   {\bf A}_n\wedge\hat{\bf k}_n \cos({\bf k}_n\cdot{\bf x} + 
\omega_n t) & + \nonumber \\
   {\bf B}_n\wedge\hat{\bf k}_n \sin({\bf k}_n\cdot{\bf x} + 
\omega_n t) &
\label{KS}
\end{eqnarray}
where $N_k$ (typically of order 100) is the number of modes,
$\hat{\bf k}_n$ is a random unit vector (${\bf k}_n = k_n \hat{\bf k}_n$),
and the directions and orientations of ${\bf A}_n$ and ${\bf B}_n$ are
chosen randomly under the constraint that they be normal to $\hat{\bf k}_n$
and uncorrelated with the directions and orientations of all other wave
modes. Note that the velocity field (\ref{KS}) is incompressible by
construction, and also statistically stationary, homogeneous and
isotropic as shown by \cite{FHMP92} and \cite{FV98}.
The amplitudes $A_n$ and $ B_n$ of the vectors ${\bf A}_n$ and ${\bf B}_n$
are determined by $ A_n^2 = B_n^2 = \frac{2}{3} E(k_n) \Delta k_n $
where $ E(k) $ is the energy spectrum prescribed to be of the form
\begin{equation}
   E(k) = \frac{ u'^{2}(p-1) }{ 2(L/2\pi)^{p-1} } \, k^{-p}
\end{equation}
in the range $ 2\pi/L_1 = k_1 \le k \le k_{N_k} = 2\pi/\eta $,
and $E(k)$=0 otherwise; $ u' $ is the rms velocity of the KS
turbulent-like flow; $ \Delta k_n = (k_{n+1} - k_{n-1})/2 $ for
$ 2 \le n \le N_k -  1 $, $ \Delta k_1 = (k_2 - k_1)/2 $ and
$ \Delta k_{N_k} = (k_{N_k} - k_{N_k-1})/2 $.
The distribution of wavenumbers is geometric (see Flohr \& Vassilicos
2000), specifically $ k_n = k_1 a^{n-1} $ with a constant $a$
determined by $ L/\eta = a^{N_k -1} $. The frequencies $\omega_n$
in (\ref{KS}) are proportional to the eddy-turnover frequency of
mode  $n$, i.e. $\omega_n = 0.5\sqrt{k_n^3 E(k_n)}$.

We have varied the power $p$ of the energy spectrum in the range
$ 1<p<3 $ without changing $u'$. 
For a given value of $p$, we calculate the number of stagnation points
by the Newton-Raphson method for different values of $L/\eta$. It turns
out, as in the case of DNS turbulence, that the vast majority of stagnation
points are straining stagnation points irrespective of the values of
$p$ and $L/\eta$ (see figure \ref{fig1}). We also find that, when
$p=5/3$, $ n_s $ is proportional to $ (L/\eta)^{D_s} $ with $D_s=2$ (see
figure \ref{ns_KS}), in agreement with our DNS and wind tunnel results.
The relation (\ref{Davilicos}) is found to hold for all values of $p$
between 1 and 3 in instantaneous realisations of our KS field and in
fact $D_S$ decreases towards 0 as $p$ increases towards 3. Varying $p$
in KS is therefore a good way to tamper with the fractal structure of
the set of straining stagnation points in the turbulent-like flow and
study what the effects of this tampering are on turbulent pair-diffusion.

\begin{figure}
\includegraphics[width=8cm]{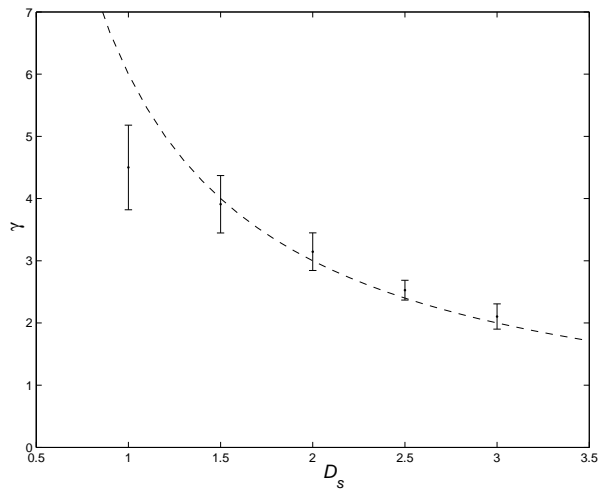}
\caption{Relation between Richardson's pair diffusion exponent $\gamma$
       and the fractal dimension of stagnation points $D_s$. The values
       of $\gamma$ are obtained from linear fits of the time dependence
   	 of $ \overline{\Delta^{2}} $ (in log-log scale) over the interval
		 where $(3\eta)^2 < \overline{\Delta^{2}} < (L/3)^2$. The length of
		 the error bars is twice the r.m.s. of $\gamma$ within that interval.
		 In these KS calculations, $N_k=40$, $L/\eta=10^3$ and the initial
		 separation is $\eta/2$ for 2000 different particle pairs. The
       dashed line shows (\ref{gammaD}) for comparison.\label{fig_gD}}
\end{figure}

We have calculated the time-dependence of the mean square distance
between pairs of fluid elements in KS turbulent-like flows for
different values of $p$ between 1 and 2 and have found that
(\ref{Richardson}) is valid in the inertial range with
\vspace*{-9mm}

\begin{equation}
   \gamma = 6/D_s 
\label{gammaD}
\end{equation}
(see figure \ref{fig_gD}), and that the Richardson constant $G$  is an
increasing function of $D_s$. For values  of $p$ between 2 and 3 we do
not find  evidence of a well-defined power  law (\ref{Richardson}) and
of course  nothing like (\ref{gammaD}).
The reason, which we believe lies behind this pair-diffusion behaviour, is
that for $p$ between 2 and 3 the power spectrum $k^{2}E(k)$ of the strain
rate field is concentrated at the smaller wavenumbers when $p>2$ but
increases with wavenumber when $p<2$. The energy spectrum of a homogeneous
and isotropic gaussian velocity field such as (\ref{KS}) for large enough
$N_k$ is related to the fractal dimension of zero-crossing surfaces by
Orey's relation $p+2(D-2)=3$ (see \cite{O70}) which implies $ p+2D_s/3=3 $.
\cite{FV98} have shown that, as a consequence of pair-diffusion locality,
$ \gamma = 4 / (3-p)$. Hence, (\ref{gammaD}) is effectively a consequence
of velocity field gaussianity and pair-diffusion locality.

\begin{figure}
\includegraphics[width=8cm]{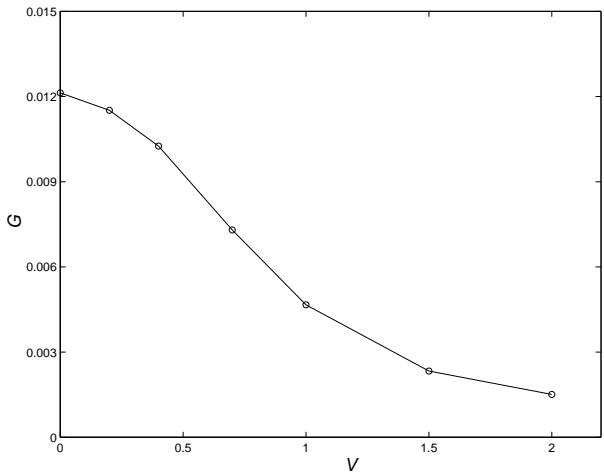}
\caption{Richardson constant $G$ versus flow average velocity $V$ using
         the parameters of the simulations as in figure \ref{fig_gD}.
       \label{G_vs_V}}
\end{figure}

Another way to modify the straining flow structure of the KS
turbulence is by adding a constant (time and space independent)
velocity vector ${\bf V}$ to the KS velocity field (\ref{KS}). The KS
velocity field defined in (\ref{KS}) is a mean-zero velocity field and
the addition of ${\bf V}$ amounts to a superposition of a constant
mean flow. The aim is to look for stagnation points of $ {\bf u}+{\bf
V} $ and monitor the changes in the straining stagnation point
structure caused by changes in $ V = |{\bf V}| $ whilst keeping $p$
constant. Our first observation is that (\ref{Richardson}) remains
valid with the same value of $D_s$ but $C_s$ decreases with $V$ (see
figure \ref{ns_KS}). By releasing fluid element pairs in the velocity
field $ {\bf u}+{\bf V} $ we can study modifications to the Richardson
law (\ref{Richardson}) caused by the mean flow velocity ${\bf V}$. The
effects on (\ref{Richardson}) parallel those on (\ref{Davilicos}): the
exponent $\gamma$ remains well-defined in the same range of times and
is independent of $V$ but the Richardson constant $G$ decreases with
increasing $V$ (figure \ref{G_vs_V}).  The addition of a constant mean
flow velocity leaves the scaling exponents $\gamma$ and $D_s$ of the
Richardson law and of the fractal straining stagnation point structure
intact, but reduces the overall number of straining stagnation points
per unit volume and also the overall extent of turbulent pair
diffusion. In view of this conclusion and also of relation
(\ref{gammaD}) there is clearly a correlation between Richardson pair
diffusion and the fractal spatial distribution of straining stagnation
points in the turbulent-like flow.  Such a correlation hints at a
certain persistence in time of the streamline structure of the flow
which causes pairs to separate when they meet straining stagnation
points. This is consistent with our results that $G$ is an increasing
function of $D_s$ and of $1/V$, i.e. an increasing function of the
number of straining stagnation points in both cases.

\begin{acknowledgments}
We are grateful to Mr. Daniel Polo and Dr. Peter Flohr for help with
laboratory and DNS data respectively. JCV acknowledges support from the
Royal Society of London and the Hong Kong Research Grant Council (project
number HKUST6121/00P).
\end{acknowledgments}

\end{document}